\def\year{2018}\relax
\documentclass[letterpaper]{article} 
\usepackage{aaai18}  
\usepackage{times}  
\usepackage{helvet}  
\usepackage{courier}  
\usepackage{url}  
\usepackage{graphicx}  
\usepackage{balance}
\usepackage{multirow}
\usepackage{multicol}

\usepackage{booktabs} 

\usepackage{multirow}
\usepackage{float}
\usepackage[normalem]{ulem}
\useunder{\uline}{\ul}{}
\usepackage{subfig}
\usepackage{subfloat}
\usepackage{graphicx}
\usepackage{amssymb}
\usepackage{xcolor}
\usepackage{paralist}
\usepackage{gensymb}

\begin{document}

%

\title{Analyzing and characterizing political discussions in WhatsApp public groups}

\author{Josemar Alves Caetano, Jaqueline Faria de Oliveira, H\'elder Seixas Lima, Humberto T. Marques-Neto\\ 
Pontif\'icia Universidade Cat\'olica de Minas Gerais (PUC Minas) 
\\Belo Horizonte, MG, Brazil \\
\{josemar.caetano,jaqueline.oliveira,helder.lima\}@sga.pucminas.br, humberto@pucminas.br \\
\AND Gabriel Magno, Wagner Meira Jr, Virg\'ilio A. F. Almeida \\
Universidade Federal de Minas Gerais (UFMG) 
\\Belo Horizonte, MG, Brazil\\
\{magno,meira,virgilio\}@dcc.ufmg.br
}

\maketitle

\begin{abstract}
\begin{quote}
We present a thorough characterization of what we believe to be the first significant analysis of the behavior of groups in WhatsApp in the scientific literature. Our characterization of over 270,000 messages and about 7,000 users spanning a 28-day period is done at three different layers. 
The message layer focuses on individual messages, each of which is the result of specific posts performed by a user. 
The user layer characterizes the user actions while interacting with a group. 
The group layer characterizes the aggregate message patterns of all users that participates in a group. 
We analyze 81 public groups in WhatsApp and classify them into two categories, political and non-political groups according to keywords associated with each group. Our contributions are two-fold. First, we introduce a framework and a number of metrics to characterize the behavior of communication groups in mobile messaging systems such as WhatsApp. Second, our analysis underscores a Zipf-like profile for user messages in political groups. Also, our analysis reveals that Whatsapp messages are multimedia, with a combination of different forms of content. Multimedia content (i.e., audio, image, and video) and emojis are present in 20\% and 11.2\% of all messages respectively. Political groups use more text messages than non-political groups. Second, we characterize novel features that represent the behavior of a public group, with multiple conversational turns between key members, with the participation of other members of the group. 
\end{quote}
\end{abstract}

\section{Introduction}
\label{sec:introduction} 
WhatsApp is a distinct and rapidly growing component of the global information and communication infrastructure. It is a private and closed  instant messaging system with  more than 1.3 billion users. WhatsApp combines one-to-one, one-to-many, and group communication by offering private chats, broadcasts, and private and public group chats~\cite{seufert2016group}, where people can share documents, photos, and videos. The use of WhatsApp for news is becoming popular all over the world, for it is more private and it does not filter content algorithmically as many social media platforms do. The Digital News Report 2017 by the Reuters~\cite{newman2017reuters} shows that significant fractions of the population of Malaysia (51\%), Brazil (46\%), Chile (39\%),  Singapore (38\%), Hong Kong (36\%), and  Spain (32\%)  use WhatsApp to find, share or discuss news in a given week. Large parts of the population in different countries of Asia, Latin America and Africa are users of the mobile messaging platform. Surveys~\cite{statistica} show that India has more than 200 million users and  Brazil has more than 120 million users.   Due to its impressive penetration in many  countries, WhatsApp  turned out to be an important platform for political propaganda  and election campaigns. In recent elections in countries such as UK, Israel, India, Spain, and Australia,  politicians have  turned to WhatsApp in an attempt to persuade voters.  But, despite its global popularity and importance, WhatsApp is notoriously opaque.  
Given the prominence and continued growth of instant messaging apps, it is natural to ask whether their characteristics are similar to those of more traditional components of the global communication infrastructure, such as Skype, Gmail, Twitter, or Facebook. Indeed, over the last few years, there have been a very limited number of studies that explored the various aspects of mobile messaging apps~\cite{church2013s,huang2015fine,seufert2016group,guo2018answering}.  Such studies are important because they allow us to understand how WhatsApp and other instant messaging systems impact the social, political and economic life of different sectors of society in different parts of the world.  

In this paper we focus on the analysis of the behavior of group chats in WhatsApp. Although being a closed system with private chats, it has a class of  public groups that  allow members to share invites on  public platforms, thus bringing more people to participate of different  discussion groups.   This paper provides  a characterization for a unique data set capturing thousands of users interacting over 81 public groups.  While an interesting subject on its own, the characterization of group chats is likely to be of paramount importance given the increasing role of WhatsApp as a major global communication  channel.   To understand the behavior of WhatsApp groups, we define a hierarchical approach to characterize the data collected from different groups. The three layers of the hierarchy are: message, user, and group. Within each layer, an analysis of different measurements is conducted. The approach is illustrated by presenting a detailed characterization of two types of thematic groups: political and non-political.  

\begin{figure}[t]
\centering
\includegraphics[scale=0.37]{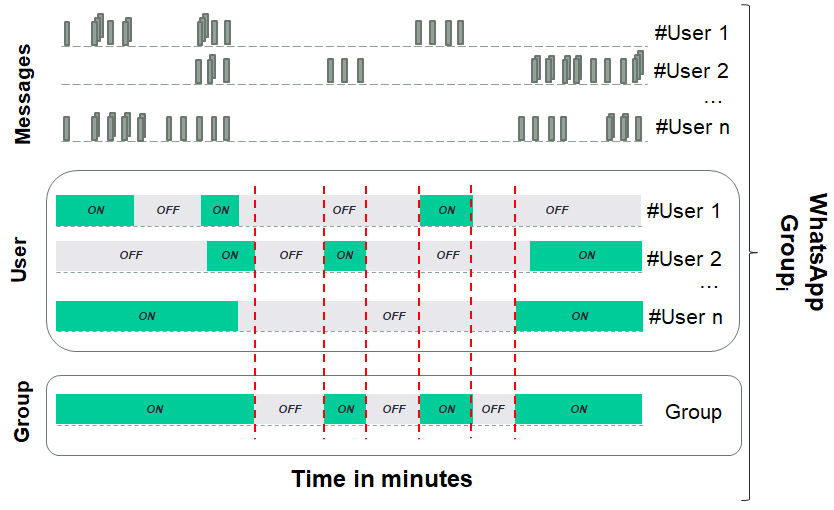}
\caption{Relationship between user messages and ON/OFF times at the group
and user session layers}
\label{fig:on_off_session}
\end{figure}

Figure \ref{fig:on_off_session} shows the hierarchical framework that captures the dynamics of WhatsApp groups. User activities, i.e., message posts, are mapped onto various session ON and OFF times. In particular, a session is the period of time during which the user is continually interacting with the group. During a session, the period of inactivity of a user does not exceed a preset threshold. A group session time is defined by the overlap of user session times. Our approach is to analyze each layer individually in order to obtain a characterization of the message arrival process and temporal statistics. Our analysis covers properties such as: session inter-arrival times, group session duration, user ranking, message distribution per session, and number of active sessions.  

The remainder of this paper is organized as follows. In Section 2, we give a high-level description of the data collection process and the data sets used in our WhatsApp characterization. This is done from three different projections. The first characterizes the aggregate message patterns of all users of a group. This is the message characterization, which we present in Section 3. The second characterizes how individual users interact within the group. This is the  user characterization, which we present in Section 4. The third characterizes the group behavior,  which we present in Section 5.  We present the different roles users play in a given group, which we describe in section 6.  We put our work in context by reviewing related research in Section 7, and we conclude with a summary of findings and of current and future research in Section 8.

\section{Collecting and Organizing \\ WhatsApp Dataset}



In this section, we present a process to collect data from WhatsApp public groups and explain how these groups are classified as \textit{political} and \textit{non-political}. We also provide  an overview of the dataset and discuss WhatsApp privacy policies and restrictions.

\subsection{Data Collection Process}

\begin{figure}[t]
\centering
\includegraphics[scale=0.35]{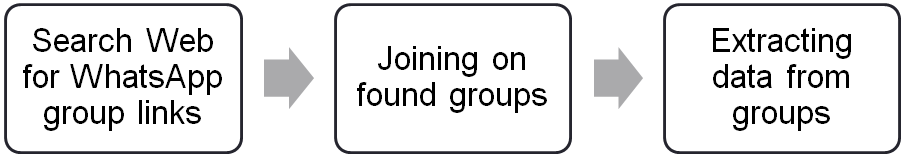}
\caption{The data collection process.}
\label{fig:data_collection}
\end{figure}

Figure \ref{fig:data_collection} shows the proposed data collection process: (i) to search for WhatsApp group links on the Web, (ii) to join these groups and (iii) to extract data from them as a group member. We use the Selenium tool (seleniumhq.org) to perform operations in  WhatsApp Web 
as a member, i.e., clicking, scrolling up and down, and obtaining the messages and other group metadata such as its name, creation date and time, and the list of all members. Next, we describe each step of this process.

\textbf{Search Web for WhatsApp group links:} First, we define a list of keywords to be searched together with ``chat.whatsapp.com'' (part of the link shared by  administrators of a WhatsApp group). We use Selenium scripts on Google Search website to get the links of WhatsApp public groups and verify whether each one points to a valid WhatsApp group. Thus, we obtain a set of WhatsApp public groups, including the ones that are published on sites which aggregate these kind of links, like 
https://grupos-whatsapp-brasil.com.br/.


In order to build the corpus for this work, we create a list of words (in Brazilian Portuguese) about some controversial topics such as politics, social movements, 2018 Brazil Presidential Election, religion, racism, misogyny, sexuality, violence, and discrimination. This \textit{bag of words} is available on https://goo.gl/cXDiFF. Using these keywords we collect 21,350 WhatsApp public groups, from which 13,525 are valid, and 7,825 are broken links.

\textbf{Join groups found:} our script simulates the clicking on a valid group link and confirming the public invitation of joining automatically a WhatsApp public group. We select the groups which have on its name at least one keyword of the selected \textit{bag of words} and have at least 64 members (25\% of 256, the maximum number allowed). As we need a valid cellphone number, we use one smartphone for each WhatsApp account. We also use computers with a Google Chrome browser to access each account through WhatsApp Web, which is the desktop platform for Whatsapp. Finally, we manually read the QR-Code required for accessing the WhatsApp Web for the first time on each computer.

We verify the name of the 13,525 valid groups and find 773 having at least one keyword from our \textit{bag of words}. Due to time limitations and the number of cellphone numbers available to create WhatsApp accounts for data collection, we join just 127 of the 773 public groups, from which 81 have at least 64 members. Thus, in total, we join 81 public groups using three Android smartphones (27 groups on each one) and employed three computers with Google Chrome to use WhatsApp Web for collecting group data.





\textbf{Collect data from each group:} in this step, our script connects to one of the WhatsApp Web accounts and opens each public group the account is connected to. A member can only access the data of the group after she joins on it.  We then  collect the group's metadata (name, creation date/time, and list of members with username and phone number) and the messages (text, image, video, audio, document, contact, location, and emojis) that were posted after the date the user joined the group. Our script anonymizes the sensitive collected data by associating a unique ID for each user. We only kept the cellphone country code of the administrators to validate, in conjunction with the main language used in the messages, the country where most of the activity of the group happens.

In order to mitigate the temporal bias, we choose to collect messages and metadata using the same period of time for all the groups. We collect 273,468 messages of 6,967 users engaged in 81 WhatsApp public groups spanning a 28-day period.


\subsection{Defining Political and Non-political Groups}
We manually classify the 81 WhatsApp public groups into two categories: \textit{political groups} and \textit{non-political groups}, based on their names and the topics identified by the LDA analysis \cite{Blei03latentdirichlet} of the  messages. When a group has neither ``political words'' on its name nor political topics associated with it, we consider it a non-political public group. Otherwise, we consider it a political public group. In total, we identified 49 political groups and 32 non-political groups.


\subsection{Selecting the Top 4 Groups}

For the characterization we present in this work we selected the top 4 political and top 4 non-political groups in terms of the number of messages posted during the 28-day data collection (at least one message per day).

The top 4 political (P1, P2, P3, and P4) have LDA topics related to ``2018 Brazil Presidential Election'', mainly containing the word ``president'' and the name of the current President of Brazil (Temer) as well as the name of the most popular pre-candidates: Lula and Bolsonaro. On the other hand, the top 4 non-political groups (N1, N2, N3, and N4) discuss different topics, as detected by LDA. The main topic of N1 is ``drugs'' (main words: marijuana, drugs, and smoking), the topic of N2  is ``Internet'' (main words: access, enter, and Internet), the main topic of N3  is ``disease'' (main word: cancer),  and the topic of N4  is ``religion'' (main words: Lord, God, Jesus, Word, Christ, and prayer). All topics of the groups are obtained by analyzing the corresponding messages with an LDA tool.

\begin{table}[t]
\centering
\caption{Number of Users and Messages of Groups}
\label{table:basic_statistics_groups}
\begin{tabular}{|c|r|r|r|}
\hline
\textbf{Group} & \multicolumn{1}{c|}{\textbf{\begin{tabular}[c]{@{}c@{}}Active \\ Users\end{tabular}}} & \multicolumn{1}{c|}{\textbf{\begin{tabular}[c]{@{}c@{}}Passive \\ Users\end{tabular}}} & \multicolumn{1}{c|}{\textbf{Messages}} \\ \hline
\hline
\hline
\textit{All groups (81)} & \multicolumn{2}{|c|}{\textbf{\textit{6,967}}} & \textbf{\textit{273,468}} \\ \hline
\hline
\textit{Both Top 4 (8)} & \textbf{1,289} & \textbf{979} & \textbf{69,914} \\ \hline
\hline
\textit{Non-Political (4)} & \textbf{408} & \textbf{374} & \textbf{12,941} \\ \hline
\hline
\multicolumn{1}{|r|}{\textit{N1}} & 70 & 67  & 4,788  \\ \hline
\multicolumn{1}{|r|}{\textit{N2}} & 110 & 100 & 4,384  \\ \hline
\multicolumn{1}{|r|}{\textit{N3}} & 143 & 139 & 2,071  \\ \hline
\multicolumn{1}{|r|}{\textit{N4}} & 85  & 68 & 1,698   \\ \hline
\hline
\textit{Political (4)} & \textbf{881} & \textbf{605} & \textbf{56,973} \\ \hline
\hline
\multicolumn{1}{|r|}{\textit{P1}} & 224 & 192 & 25,892  \\ \hline
\multicolumn{1}{|r|}{\textit{P2}} & 158 & 141 & 11,615  \\ \hline
\multicolumn{1}{|r|}{\textit{P3}} & 250 & 154 & 9,924   \\ \hline
\multicolumn{1}{|r|}{\textit{P4}} & 249 & 118 & 9,542   \\ \hline
\hline
\end{tabular}
\end{table}

\subsection{Dataset Overview}
\label{sec:collection:dataset}

We collect data from WhatsApp public groups from October 10, 2017 to November 06, 2017.
The data collected from the 8 selected public groups (4 political and 4 non-political) comprise 69,914 messages (25.5\% of the 273,468 messages collected overall) posted by 1,289 users (18.5\% of 6,967 users that posted anything in the dataset). The administrators of all these 8 groups use cellphones with Brazilian Area Code (55), characterizing them as Brazilian public groups.

Table \ref{table:basic_statistics_groups} presents an overview of the number of users and messages of all 81 groups as well as the groups selected for characterization. We define active users as those who post at least one message during the period analyzed. Passive users, on the other hand, did not publish any message in the group during the same period. Table~\ref{table:dataset} shows the number of each type of WhatsApp message (i.e., text, media -- audio, video, and images --, emojis, and links). Because a message may combine more than one type of media, the sum of the types is greater than the number of messages of the 8 selected groups.


\begin{table}[t]
\centering
\caption{Number of Messages per Type (only top 4 groups)}
\label{table:dataset}
\begin{tabular}{|l|r|}
\hline
Text & 50,210 \\ \hline
Media (audio, video, image) & 14,012 \\ \hline
Emoji & 7,842  \\ \hline
Links & 5,041 \\ \hline
\end{tabular}
\end{table}



\subsection{Privacy Policies and Restrictions}
\label{sec:collection:policies_and_restrictions}

Our data collection process complies with WhatsApp privacy policies, which are available on the official site (www.whatsapp.com/legal/?l=en. Accessed on Nov 24, 2017). The privacy policy states that a user should be aware that her data is shared  with other members of the group. All users in a WhatsApp group (public or not) can see the profiles of other users as well as the  messages posted in the group (see WhatsApp FAQ at faq.whatsapp.com). It is possible to extract the messages from the group and send them by e-mail or share them with other users or groups.


Although WhatsApp enables users to save the messages and send them by e-mail, it is possible to send just up to the 10,000 most recent messages with media or 40,000 without media (faq.whatsapp.com). WhatsApp users may  save messages from private chats or groups (public or not). We use  a data collection process to automate the extracting data from WhatsApp Web application that follows its privacy policy.

\section{Hierarchical Characterization of  WhatsApp Public Groups}


In this section we describe the  hierarchical characterization~\cite{Veloso:2006:HCL:1133553.1133564} of  WhatsApp public groups. The characterization has three distinct layers: \textit{message}, \textit{user}, and \textit{group}, as shown in Figure~\ref{fig:on_off_session}. We analyze each layer separately considering specific metrics for each one in order to understand the user behavior in the  chat groups. 


\textbf{Message Layer:} The top layer of our hierarchy focuses on the characteristics of the messages of a group. Each message is associated with a  user, that is  identified by a unique ID created from her phone number on WhatsApp. The message characteristics we consider are: message content types (i.e.,  text, media, link), arrival time, and the number of messages posted in the same period of time. The time granularity of WhatsApp data collection is one minute.

\textbf{User Layer:} Focusing on an individual user of a group, we move to the second layer of our hierarchy and look at the statistics that characterize a  user session. We define a user session as the interval of time during which the user is actively engaged in posting messages in a  WhatsApp group, such that the duration of any period of inactivity does not exceed 15 minutes. Thus, the engagement pattern of a given user is characterized by periods of activity (session ON time) and inactivity (session OFF time). Figure \ref{fig:on_off_session} shows how user activities (i.e., posting a message) translate to ON and OFF times. 

\textbf{Group Layer:} Zooming in on session ON times, we characterize the bottom layer of our hierarchy, which focuses on group sessions, each of which is the result of interleaved user sessions. Thus, a given group session is characterized by periods of users posting messages (ON time) and of silence (OFF time). Figure \ref{fig:on_off_session} shows how users activities (start and stop discussions) result in various ON and OFF times of a group. In addition to characterizing  ON and OFF times, we also compute group sessions durations, number of concurrent user sessions, and group session inter-arrival times. 

Characterizing the workload at these distinct levels of abstraction allows one to concentrate on the analysis of the behavior of the different groups that exists in this type of application, namely political groups and non-political groups. This hierarchical characterization can also be used to capture changes in user behavior and map the effects of these changes to the lower layers of the hierarchical model, i.e., session and group layers. Finally, this layered approach enables us to perform a thorough characterization to understand how users engage themselves on political and non-political discussions on WhatsApp groups.


\section{Message Layer Characterization}

In this layer, we characterize the messages posted by users in the public groups. We analyze the temporal behavior of the messages, the average interval between messages during the periods of the day and the types of message content. 

\subsection{Temporal Behavior of Messages}

\begin{figure*}[h]
\centering
\includegraphics[width=\textwidth]{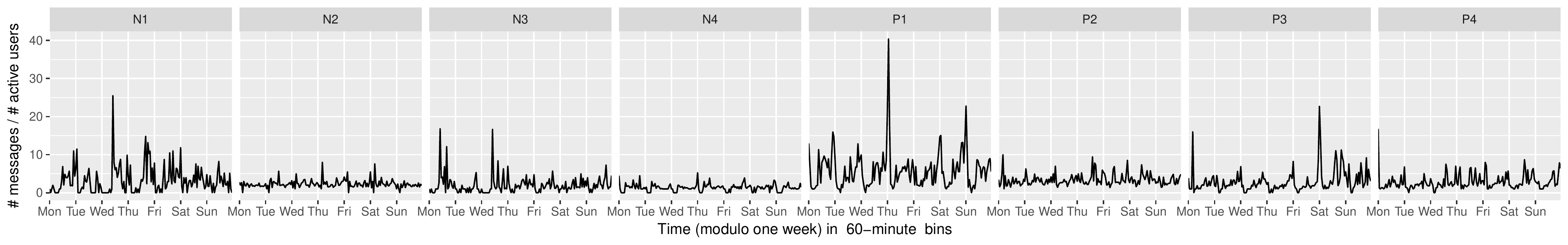}
\caption{Temporal behavior of message number published by active users over week days}
\label{fig:Ratio_messages_by_users} 
\end{figure*}

At any point in time $t$, there may be some users posting messages in the groups. The ratio $r(t)$ of the number of messages to number of active user allows us to analyze the interaction level of users over  $t$. Figure \ref{fig:Ratio_messages_by_users} shows $r(t)$ in intervals of 60 minutes each, over a weekly period. We noted that there is not a clear pattern of $r(t)$ concerning the day of the week. However, we could see that for all groups the $r(t)$ tends to decrease in early hours of the day and start increasing  in the morning. 

\begin{figure}[h]
\centering
\includegraphics[width=\columnwidth]{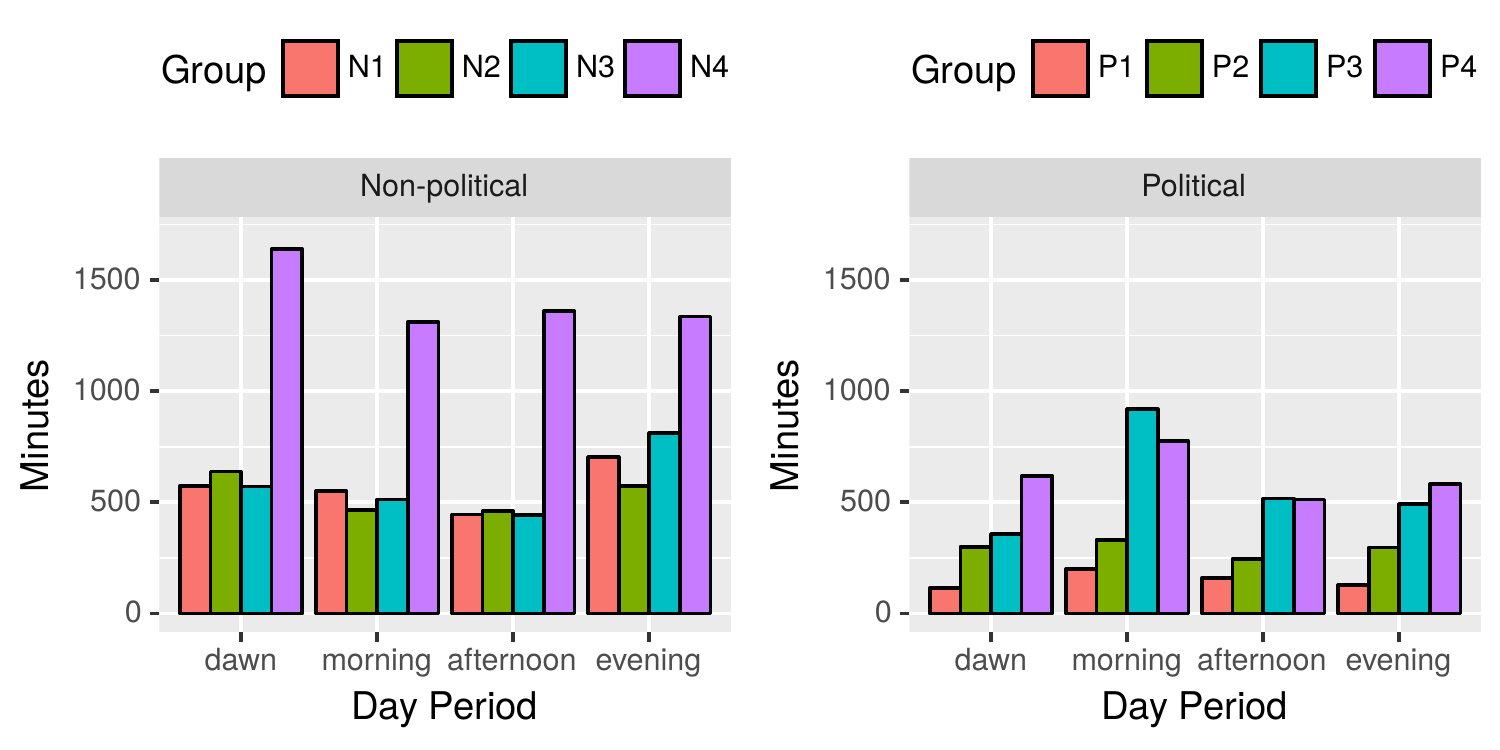}
\caption{Messages IAT throughout the day}
\label{fig:iat_day_period}
\end{figure}

\subsection{Message IATs}

Let $t(j)$ denote the arrival time of the $j\textsuperscript{th}$ message in a group. Let $a(j)=t(j + 1)-t(j)$ denote the interarrival time (IAT) between the $j\textsuperscript{th}$ and $(j + 1)\textsuperscript{th}$ messages. Figure \ref{fig:iat_day_period} shows the averages of $a(j)$ over each day period.  We note that $a(j)$ of non-political groups tend to be higher than the $a(j)$ of political groups. We also note that the $a(j)$ varies according the period of the day. For the political groups the highest $a(j)$ average is over the morning, and for the non-political groups the highest $a(i)$ average is over the evening (N1 and N3) or in the early hours of the day (N2 and N4).

\subsection{Content Type}

\begin{figure}[h]
\centering
\includegraphics[width=\columnwidth]{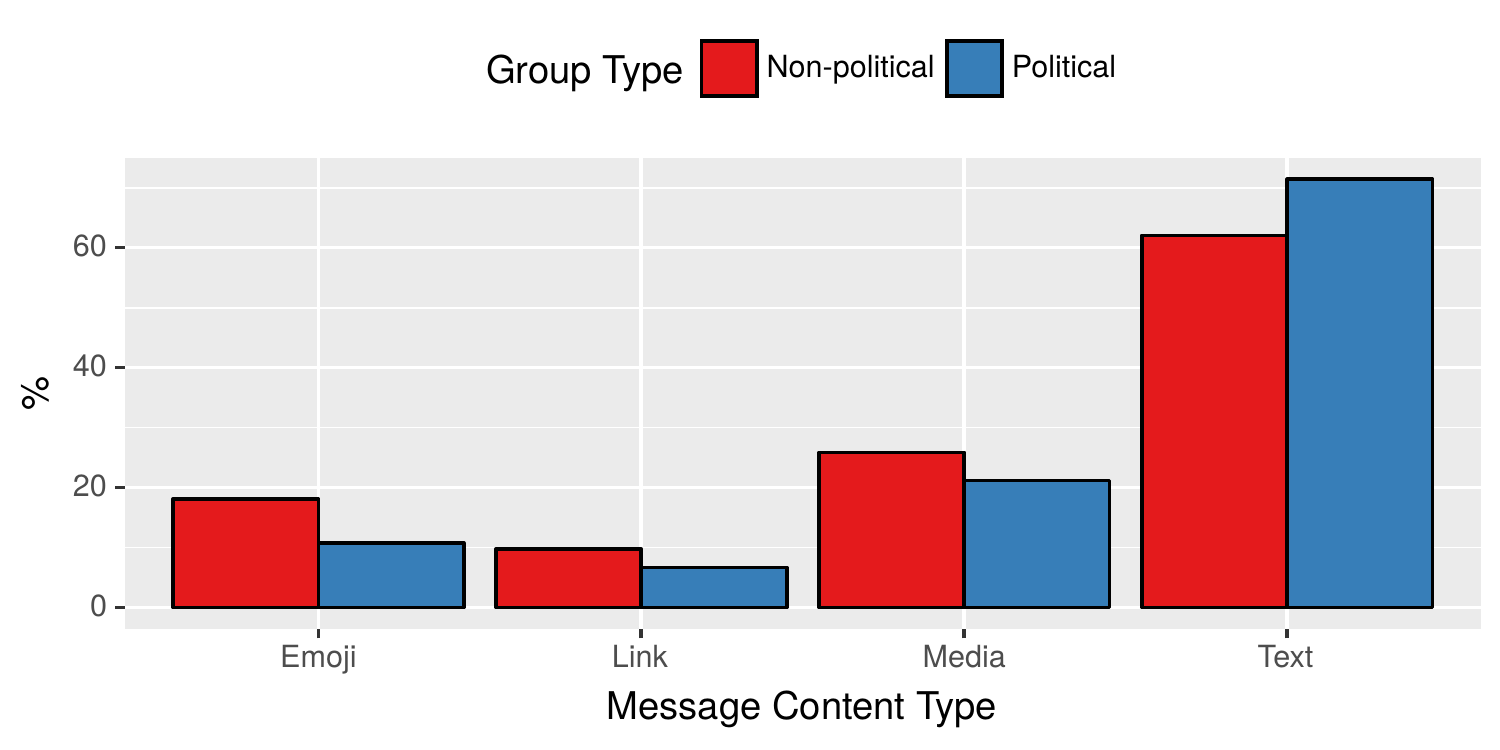}
\caption{Percentage of the message content types}
\label{fig:Percentage_messages_type} 
\end{figure}

Figure \ref{fig:Percentage_messages_type} shows the percentage of content types used in the messages. We note that messages with text are the most common message type. However, the number of messages with media content  is also significant, since at least 20\% of messages contain some image, audio or video. Users of political groups post more messages containing text than  users of non-political groups. This finding may indicate that there are more discussion among  users of political groups than among uses of non-political groups.

\section{User Layer Characterization}

In this section we focus on the User Layer of the hierarchy. We analyze the temporal characteristics by measuring the IAT and $T_{off}$ of the messages, as well as defining the threshold for defining a user session. Then, we measure characteristics of both the user itself and the user sessions. Finally, we examine the rank of the users in relation to their activity. It is important to notice that our measurements focus on active users only.

\subsection{IAT of User Messages}

For this analysis we calculate the IAT (Inter-Arrival Time) of the messages of the users. The average difference in time between messages of the same user from non-political groups is 447 minutes ($\sim$7.4 hours), while for users from political groups is 216 minutes ($\sim$3.6 hours). If we look at the CDF distribution (presented in Figure~\ref{fig:timedelta_user-per-group-type}), we can see that a considerable portion of the IATs are higher than 1 hour: $\sim$39\% for non-political groups and $\sim$21\%. These results corroborate with the fact that political groups have more users, which increases the probability of having a higher quantity of messages in the same period of time, generating lower IATs.

We choose a threshold of 15 minutes for creating the user sessions, meaning that if there's a period of 15 minutes or more, the last session will be finished, and the next message will be part of a new session. 

We measure the $T_{off}$, which is the ``waiting'' time between sessions. The user sessions of users from political groups are, in general, lower than those from the users of non-political users: 1,011 minutes ($\sim$17 hours), compared to 1,307 minutes ($\sim$22 hours). It is important to notice that, although the average being high, the standard deviation is also high (2,446 and 2,837 minutes, respectively), meaning that there are also users which are very active (having lower $T_{off}$ values). We present the CDF in Figure~\ref{fig:timedelta_user-per-group-type}. By definition, the curve starts at 15 minutes, since it is the chosen threshold for creating the user session. 



\begin{figure*}[t]
\centering
\includegraphics[width=\textwidth]{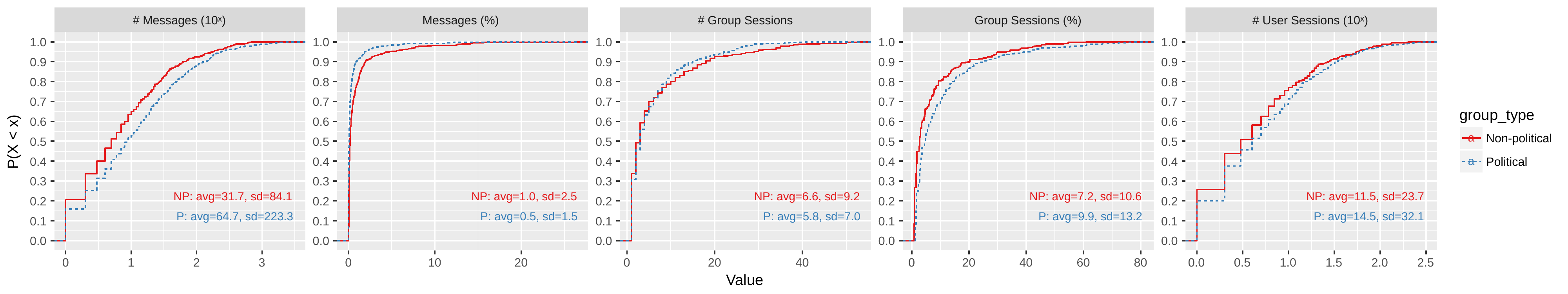}
\caption{Distribution of User Metrics.}
\label{fig:dist_user_metrics}
\end{figure*}

\begin{figure}[t]
\centering
\includegraphics[width=\columnwidth]{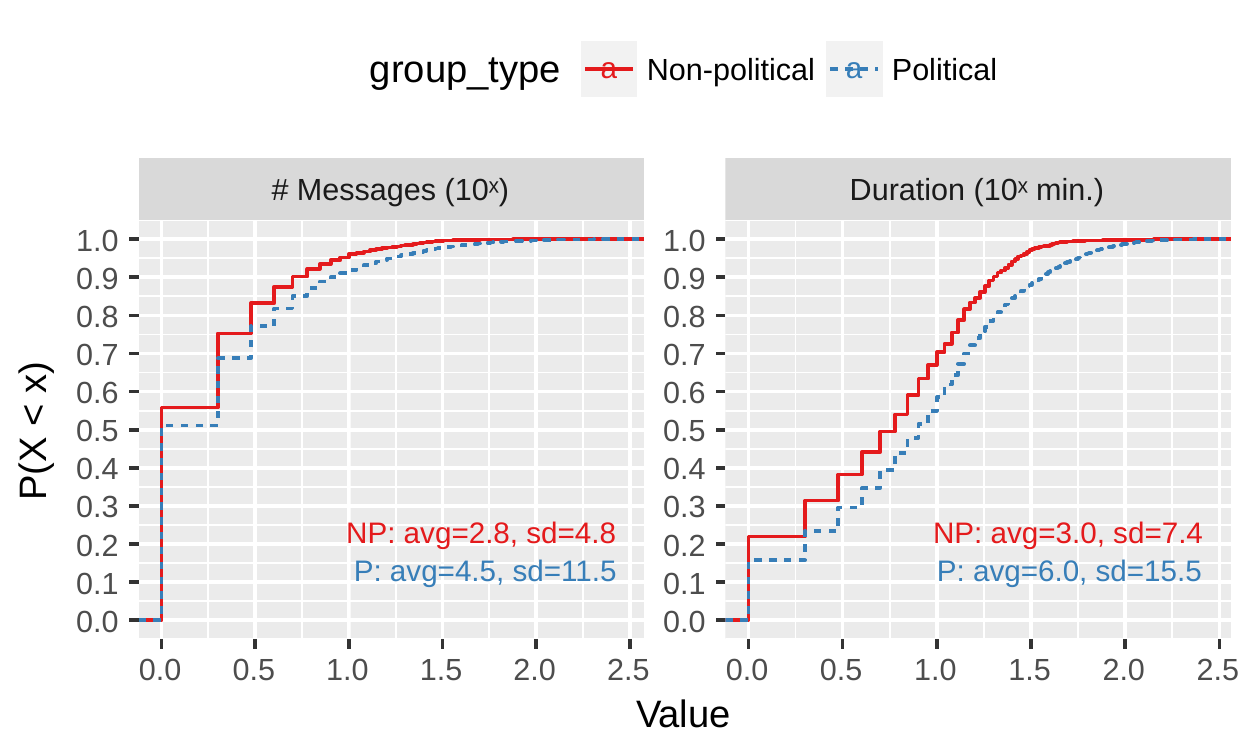}
\caption{Distribution of User Session Metrics.}
\label{fig:dist_user_session_metrics}
\end{figure}

\begin{figure*}[t]
\centering
\includegraphics[width=\textwidth]{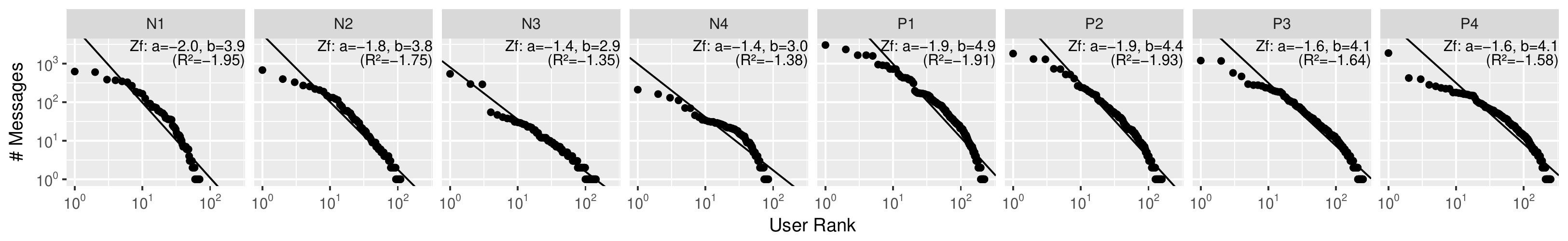}
\caption{User frequency distribution and the corresponding estimated Zipf parameter}
\label{fig:user_frequency}
\end{figure*}

\begin{figure}[h]
\centering
\includegraphics[width=\columnwidth]{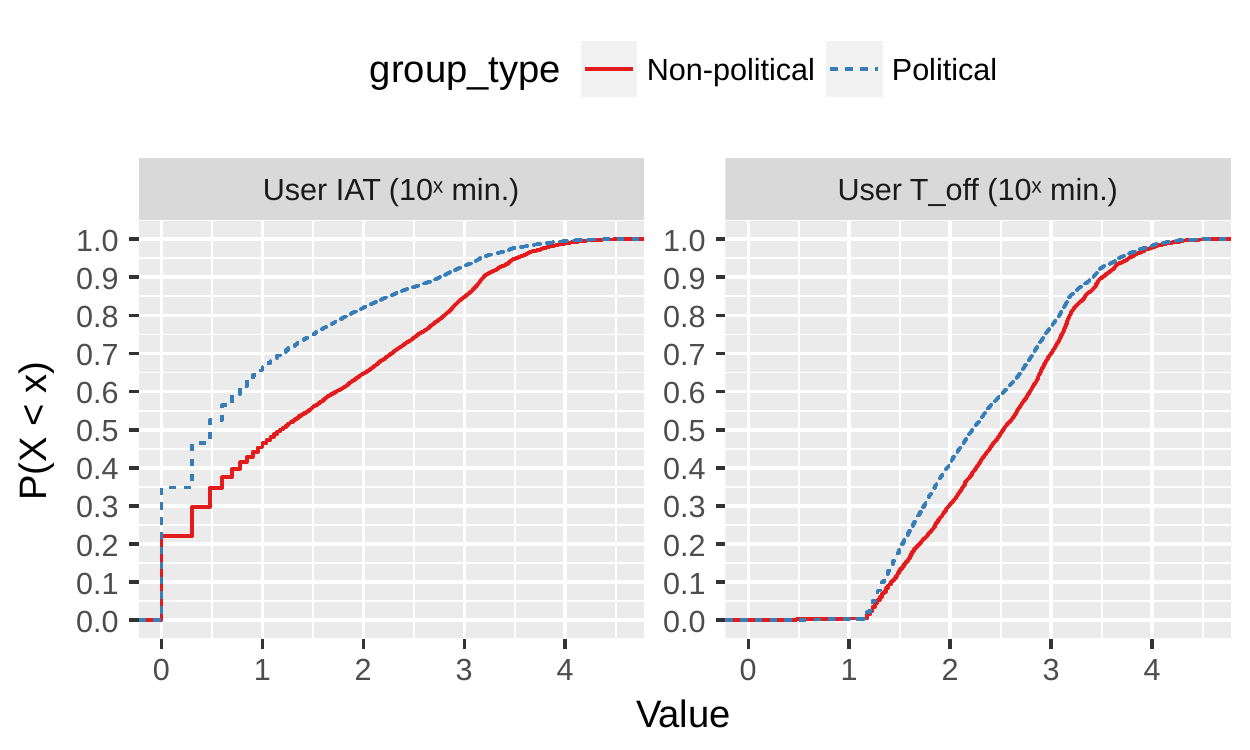}
\caption{Distribution of IAT between \textit{users sessions}}
\label{fig:timedelta_user-per-group-type}
\end{figure}


\subsection{User Metrics}

We calculate five metrics per user: (1) number of messages, (2) percentage of the messages in relation to all messages, (3) number of group sessions the user participate, (4) percentage of group sessions the user participates in relation to all, and (5) number of user sessions. We show the empirical cumulative distribution function (CDF) of the metrics, segmenting the analysis for each group and for each group type (political and non-political), presented in Figure~\ref{fig:dist_user_metrics}. 

We observe that most of the users send few messages, but users from political groups are slightly more active than those in the non-political groups. For instance, 65\% of the non-political-group users sent 10 messages or less, while only $\sim$52\% of the political-group users did the same. On the other hand, there are users which are very active: nearly 10\% of the users sends more than 100 messages. In terms of the percentage of messages, the numbers are very low for most users. Nearly 90\% of the users contribute with less than 2\% of the messages, even though having exceptions where the user can account for over 25\% of the messages in the group.

Increasing the grain of analysis, and looking at the participation of the users in group sessions, we observe that there is a difference on the distribution for each group, but the general trend for both group types is very similar. Nearly 80\% of the users participate in 10 sessions or less. If we analyze the percentage of participation, around 90\% of the users participate in only 20\% of the sessions, having some more active users engaging in up to 80\% of them.

We also count the number of individual sessions each user have, considering our previously defined interval of hiatus for ending and starting a new session. Although some users have more than 300 sessions, 90\% have 30 or less sessions.

These results of user characterization indicate that, despite the huge activity in terms of number of messages in some groups, the participation is not homogeneous. Most users act more like listeners/readers than writers, and most of the content is posted by few users that seem to dominate the discussion quantitatively. 

\subsection {User Session Metrics}

We focus now on analyzing the user sessions that are created by our methodology. We present the empirical CDF distributions in order to compare the behavior between group types. The metrics calculated are: (1) number of messages, and (2) duration in minutes. 

We observe that most of the user sessions have only one message: $\sim$56\% for non-political groups and $\sim$51\% for political ones. Actually, few user sessions have more than 10 messages: only $\sim$4\% for non-political and $\sim$8\% for political groups. Again, there are outliers, and some user sessions have as high as 284 messages. 

Now we analyze the duration of the user sessions. Similarly to the duration of group sessions, user sessions of political groups are, in general, longer than the non-political ones. For example, 42\% of the political groups' user session last more than 10 minutes, and only $\sim$30\% of them for the non-political user sessions.

\subsection {User Rank}

In this section, we inspect  individual users and their activity in the group in terms of the number of messages. We calculate the number of messages each user sent in the groups, sort the users, and give a rank number for each one. Next, we plot a scatter plot (each point is a user) where the X axis is the rank and the Y axis is the frequency (number of messages sent). We also run a Zipf's law linear fit in the log-log scale, which will investigate whether the number of messages of a user is inversely proportional to its rank. The plot for each group is presented in Figure~\ref{fig:user_frequency}, including the Zipf linear fit with the respective linear parameters (`a' is the slope and `b' is the intercept) and the $R^2$ of the regression. 

It is interesting to notice that there are indeed ``dominant'' users that send most of the messages (over 1,000 messages), while there are lots of users issuing a very low number of messages. The Zipf law regression, even though not being perfect, had a good fit for most groups ($R^2$ between 0.85 and 0.97).

\section {Group Layer Characterization}

In this section we study the Group Layer of the hierarchy. We first explain how we create the sessions, then we characterize the temporal aspects (IAT and $T_{off}$) and then me measure several metrics for each one of the identified group sessions.


\subsection {Number of Sessions}

Since each message  does not explicitly highlight the delimiters of a given session, the number of sessions in the trace depends on our choice of the session silence period parameter \cite{Veloso:2006:HCL:1133553.1133564}. To identify the ideal silence period for all groups evaluated, we develop a heuristic based on scatter plots for each group considering two variables: (i) the group's silence time and (ii) the frequency in which the times of silence are repeated in the group. We observe that there is a negative relationship between the two variables, which means that the smaller the group's silence time the greater the frequency at which the times of silence are repeated, and vice-versa. 

Therefore, the graphs present elbow curves and we define that the silence time that lies in the middle of the curve ``elbow" corresponds to the size of the group session. To identify such a point we draw a line with a 45\degree inclination from the coordinate (0,0) of the graph. The point where this line and the curve cross corresponds to the group \(T_{off}\).

To simplify our analysis we identified the average \(T_{off}\) of the top 8 WhatsApp groups (81 minutes) and defined it as the \(T_{off}\) for all groups. Table \ref{table:basic_statistics_groups} presents the basic statistics of users that posted messages (active users), users that only read messages (passive users), total sessions (considering \(T_{off}\) = 81)  minutes and total messages posted through the 28 day observation.

\begin{figure*}[h]
\centering
\includegraphics[width=\textwidth]{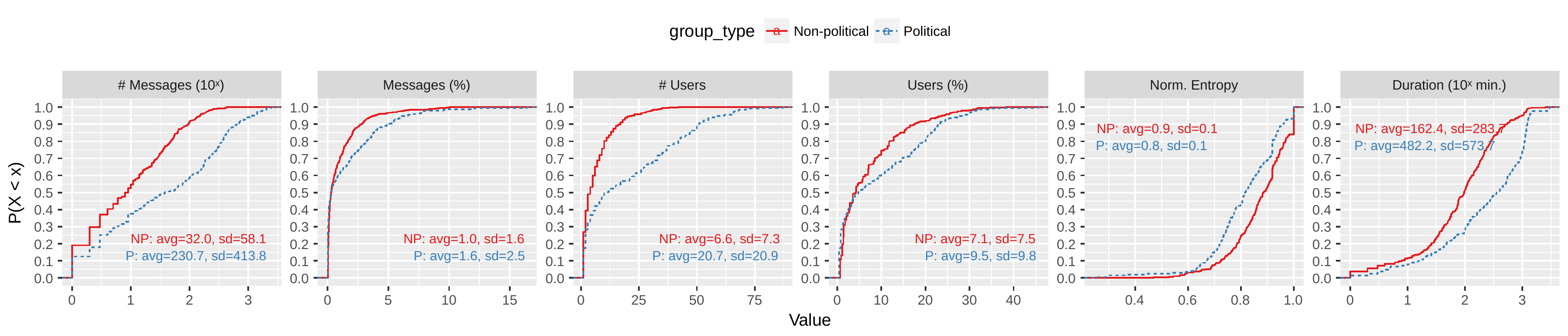}
\caption{Distribution of Group Session Metrics.}
\label{fig:dist_group_metrics}
\end{figure*}


\subsection{IAT of Group Messages}

\begin{figure}[h]
\centering
\includegraphics[width=\columnwidth]{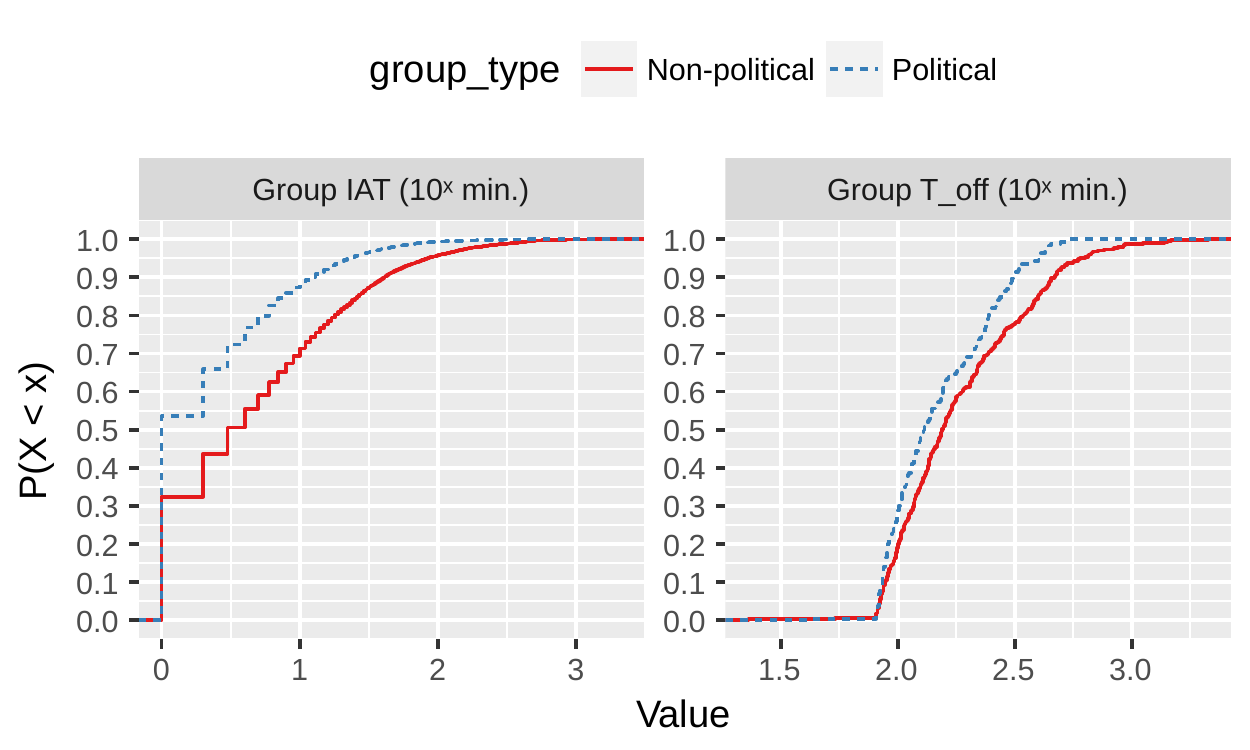}
\caption{Distribution of IAT between \textit{group sessions}}
\label{fig:timedelta_group-per-group-type}
\end{figure}

One of the aspects we want to analyze in a chat group is the time between two messages. 
To study that, we measure the IAT (Inter-Arrival Time) of the messages in a group. We focus on comparing the both types of groups in our work: political and non-political. On average, messages in political groups are separated by 2.8 minutes, while messages of non-political groups present an average IAT of 12.4 minutes. The CDF is shown in Figure~\ref{fig:timedelta_group-per-group-type}. 

Next, we analyze the time interval between two group sessions, which we call $T_off$, meaning that the amount of time with no messages, considering the group sessions defined by our methodology. Since the threshold is 81 minutes, there could not exist a $T_{off}$ below this value, otherwise it would be included in the last session. This fact is verified by the CDF presented in Figure~\ref{fig:timedelta_group-per-group-type}. As expected, the $T_{off}$ values of political groups are in general lower than the non-political ones: average of 172 minutes ($\sim$2.9 hours) and 236 minutes ($\sim$4 hours), respectively.

\subsection{Group Session Metrics}

We use CDF distributions to compare the two types of groups. The metrics we calculate are: (1) number of messages, (2) percentage of messages (i.e., coverage), (3) number of users, (4) percentage of the users among the active ones, (5) entropy and (6) duration in minutes. The plots of the distributions are presented in Figure~\ref{fig:dist_group_metrics}.

First, we analyze the number of messages in the group sessions. The distribution is very different among the groups. For instance, $\sim$72\% of the sessions of N3 had 10 messages or less, while only $\sim$20\% of P2's sessions had the same amount (i.e P2 sessions have more messages in general). There are also sessions with more than 1,000 messages. By aggregating the metric by group type, we observe that the sessions of the political groups generally comprise more messages. These results are expected, since the political groups have a higher number of participants, increasing the probability of a message being sent in the same period of time. If we now look at the coverage in terms of percentage of the messages, the difference between groups is smaller, and despite existing some sessions covering more than 15\% of the messages in the group, most of them contains less than 5\% of the messages ($\sim$90\% for political and $\sim$96\% for non-political groups).

Next, we count the number of users in each session. Similarly to the number of messages, the distribution is very different for each group, since it depends on the total number of active users (participants). Counter-intuitively, a relatively high portion of the group sessions has only one user participating ($\sim$40\% for N3). The difference between the groups can be seen, for example, considering the percentage of sessions having more than 10 users, ranging from $\sim$9\% for N3 and $\sim$72\% for P2. Again, normalizing the value and showing the user coverage in terms of percentage decreases the difference between the groups, although N3 and P2 kept having a different trend than the other groups. For instance, while 50\% of P2's sessions have 20\% or less of the users, 65\% of N3's sessions have 3\% or less of the participants engaging. Although rare, there are also some sessions engaging a high percentage of the participant, as high as 46\% in one of the groups (P2).

In order to explore the question of dominance in the discussion, we measure the entropy of the group sessions based on the number of messages each user sent. The entropy of a group session $GS$ is calculate by $H(GS) = - \sum_{u \in U_{GS}} p(u) \cdot \log_2 p(u)$, where $U_{GS}$ is the set of users participating in $GS$, and $p(u)$ is the proportion of messages of user $u$ in the group session. Since the value of entropy is sensitive to the number of users, we normalize it by dividing the value by $\log_2(|U_{GS}|)$. A higher value of entropy implies that the discussion is more evenly distributed (i.e. the users contribute in the same amount), while a lower entropy indicates the presence of dominators (i.e. few users with most of the messages). If we observe the fifth plot of Figure~\ref{fig:dist_group_metrics}, we can see that the group sessions of the political groups commonly have a lower value of entropy, meaning the presence of dominators in the discussion is stronger than in the non-political groups.

The last metric we investigate is the duration of the session or, in other words, how long the sessions last. It is possible to observe that political group sessions are generally longer than non-political ones, having 70\% of them lasting more than 100 minutes (1.6 hours), while only 50\% of the non-political. Looking for the extremes nearly $\sim$3\% of the sessions last less than a minute, and the longest session lasts 2,978 minutes (~50 hours).

\section{Analyzing Group Behavior}



\begin{figure*}[h]
\centering
\includegraphics[width=\textwidth]{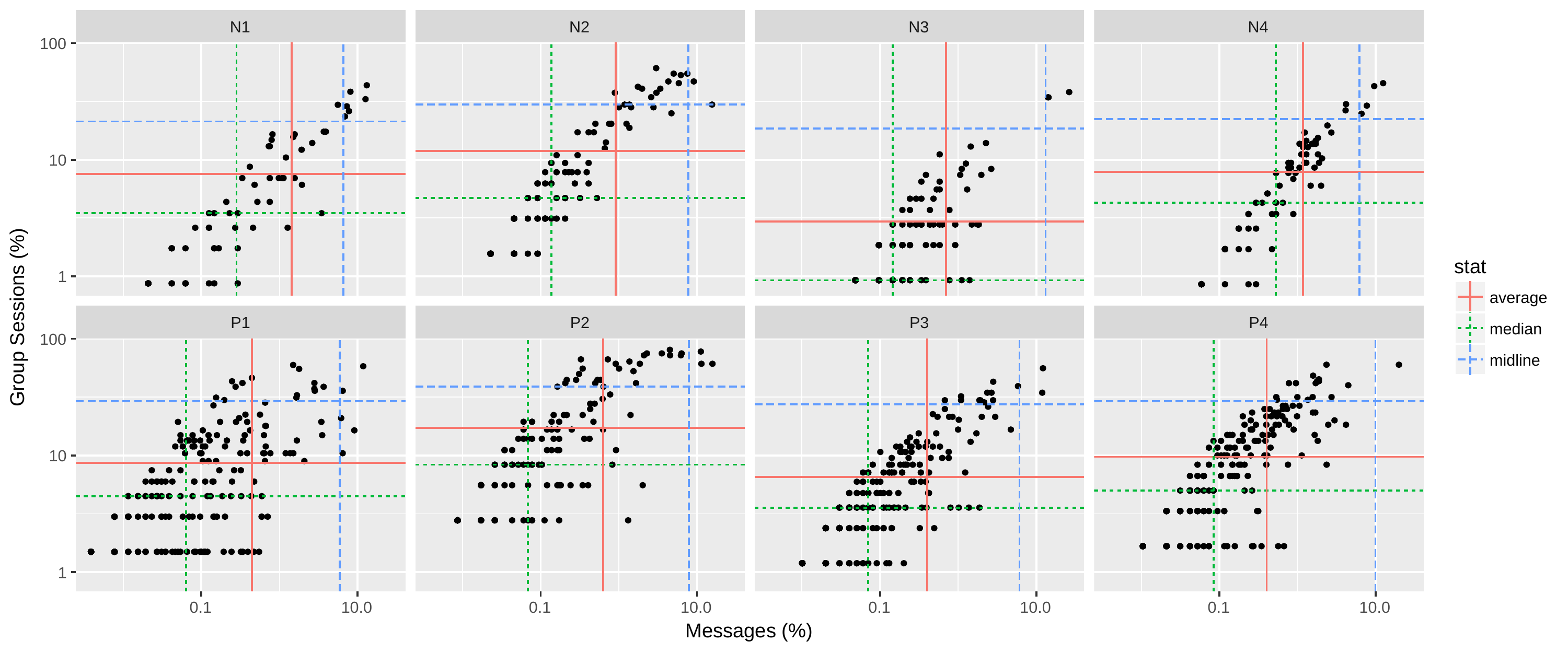}
\caption{Scatter plot of the users considering the percentage of participation in terms of messages and group sessions}
\label{fig:quadrant_msg_session}
\end{figure*}

In this section we analyze  user activity  patterns within a group chat. We only take into account the activity of users who post messages in the group.  Passive users that do not post messages in the group are also able to read messages posted by others. We call those users "audience" (or listeners), because they can  follow the discussions, although they do not place their opinion. 

The typical behavior of users  in a group is not too complex. As shown in Figure~\ref{fig:on_off_session}, a user starts a new session by posting a message. She could interact with the group by posting  one or more messages.  Afterwards, she could go out and stay silent for some time. At least two factors characterize the behavior of a user in a group: number of messages posted and the number of sessions she participates.

\paragraph{Popularity Profile} It has been well established that typical behavior in social media platforms~\cite{benevenuto2009characterizing} exhibits a significant skew in terms of the popularity of the various objects (e.g., news, songs, etc.)  accessed by users.

Thus, a natural question with respect to objects in a group is whether they exhibit a similar popularity profile. Figure~\ref{fig:user_frequency}  shows the popularity profile of users in a group in by plotting on log-log scales the frequency of posts to a group against the rank of the group. It demonstrates the use of Zipf’s Law to illustrate the presence of message concentration, sorted in decreasing order of popularity as a function of the user rank. Although the slope is not -1, there is an evidence of the message concentration property in few users. The bottom part of the figure shows that few users in political groups post more than 1,000 messages  and the majority of the users post less 100 posts in the period of analysis (i.e., 28 days). The users in the first positions of the ranking can be viewed as those that somehow lead the discussions.

A major differentiating aspect of political when compared to non-political ones is that in accessing the WhatsApp group, users are in fact engaging in an exchange of postings, which can be thought of as a dialogue between the various players – between the leaders and her audience as well as between members of the group catalyzed by a given theme. In order to characterize the attributes of this dialogue, we look at the simple  structure shown in Figure~\ref{fig:on_off_session}. In particular, such a dialogue can be seen as a sequence of postings by the participants and reading activities, that are not explicitly shown in the figure. Using these the concepts of session, we can define and characterize a number of attributes that allow us to quantify the levels of interaction induced in a given group.

\paragraph{Types of Users}  One set of attributes from Figure~\ref{fig:on_off_session} that could be used to characterize the level of user engagement to a given blog is the inter-arrival time of user sessions and  the inter-arrival time of postings. We refer to these by the inter-session, inter-posting, and inter-comment times, respectively.  Thus, sessions in a group could be seen as constituting dialogues (or sets of interactions) between group users – dialogues that are catalyzed by the themes discussed in the chats. A natural question then is whether there is a difference in the type of user behavior induced by the various patterns of posting messages.

Figure~\ref{fig:quadrant_msg_session}  shows a scatter plot in which each point represents a user. The coordinates of the user reflect the percentage of sessions a user participated (on the Y axis) and the percentage of messages (on X axis) the user posted in the group. The scatter plot shows that a correlation exists (as one would expect) between the total number of sessions  and the percentage of messages posted by users. However, the scatter plot also shows great differences among different types of users.  For instance, in political groups there are a number of users that participated of a large number of sessions, but posted few messages. They are located in the upper left quadrant of figure 12 of the political groups. They participated in more than 10\% of the sessions but posted less than 0.1\% of the messages. We do not see this type of users in non-political groups. We call those users "interested audience".

 Another  question is whether  we can find a good analogy to represent the dynamic of group behavior. A simple analogy that illustrates how a WhatsApp chat works is the model of a talk show, in in which one person (or group of people) discusses various topics put forth by a talk show host.  Usually, guests consist of a group of three to four people who  have  experience in relation to a specific issue is being discussed on the show. In addition to the host and guests there is the audience, that is composed of people that ask some questions and people that only watch/listen the show. In our analysis, we identified in the political groups some people that could be called host, because they drive the discussions by posting a large number of messages. We have people that opinionate on the issues by posting a number of messages and we have the audience, composed of passive and interested people.

\section{Related Work}


The workload characterization of computational systems has been explored in several studies. For instance, \cite{Veloso:2006:HCL:1133553.1133564} present a hierarchical characterization of live streaming media Internet workload, ~\cite{cheng2007understanding} analyze Youtube videos, ~\cite{benevenuto2009characterizing} characterize online social networks, and \cite{li2015user} characterize user behavior of large-scale mobile live streaming system. In \cite{dewes2003analysis}, the authors characterize Internet chat groups.

Although we may easily noteice the ever-increasing use of messaging applications~\cite{newman2017reuters}, there are few recent studies about these platforms.
Some works characterize WeChat by analyzing network traffic \cite{huang2015fine}, studying temporal dynamic, spatial dynamic, and user behavior \cite{liu2017patterns}, predicting invitees of social groups \cite{ijcai2017-519}, as well as predicting "reply-to" in a group chat involving multiple topics~\cite{guo2018answering}.

Talking about WhatsApp, \cite{mahajan2013forensic} focuses on conducting forensic data analysis in messaging applications, like WhatsApp, extracting data from Android smartphones using a specific hardware used on forensic context for mobile device inspections. \cite{rosler2017more} present a systematical analysis of security characteristics of WhatsApp and of two other messaging applications. They show the security vulnerabilities of the analyzed applications (including WhatsApp) by accessing, without authorization, private groups.





Furthermore, \cite{moreno2017whatsapp} collected WhatsApp messages to monitor critical events during the Ghana 2016 presidential elections. They present a tool that integrates WhatsApp to an online social media monitoring system called Aggie. However, the developed tool was able to capture only text messages, and these messages did not contain the group identification. Moreover, the tool cannot get media messages. \cite{seufert2016group} investigates the use of WhatsApp communication in group chats and, in particular, its implications on mobile network traffic through a survey. The authors also analyze the messaging history sent by survey participants. 

Our work, differently of others, presents a consistent proposal for a hierarchical characterization of user behavior on WhatsApp public groups, using a very significant data set collected in large scale. Some of our characterization results can be compared with the results found in \cite{dewes2003analysis} and \cite{seufert2016group}. If we compare the Web-chat and IRC IAT of sessions found by \cite{dewes2003analysis} with the \(T_{off}\) of our user sessions, we observe that 12\% of WhatsApp sessions have a \(T_{off}\) higher than 2 thousand seconds, while IRC is $\sim$5\% and Web-chat $\sim$0.6\%. Regarding the duration, we may compare with the duration of WhatsApp group sessions, 47\% lasts more than 100 minutes, and while for IRC and Web-chat this is around 30\%. Moreover, a small piece of our WhatsApp characterization results can also be compared with those presented by  \cite{seufert2016group}. For instance, in both work the most of the users send few messages and few users dominate sending messages. However, the huge difference in the amount of data analyzed would leave this comparison relatively vulnerable. 

\section{Concluding Remarks}

In this  paper  we  have  presented a novel characterization methodology for instantaneous messaging public groups and we instantiate it towards analyzing the similarities and differences between political and non-political Whatsapp groups. Our characterization adopts a three-layer hierarchical approach, corresponding to messages, users and groups. Our characterization results  uncover a number of interesting observations in each of these layers.

\noindent
\textbf{Message layer:} The messages from all groups present a daily temporal pattern which almost always repeats 7 days a week. The message inter-arrival time (IAT), though, is different between the political and non-political groups, since the political ones present a smaller IAT in the afternoon and across night. Regarding the content, political groups tend to post more text than the non-political groups.

\noindent
\textbf{User layer} Most of the users send few messages, but those who participate in political groups are more active. On the other hand, in all groups there are few users that are responsible for a significant number of messages. By analyzing the user sessions, we found that a significant number of sessions contain just one message, but, on average, sessions from political groups last longer than non-political ones. 

\noindent
\textbf{Group layer:} The sessions of political groups comprise more messages, last longer and have lower entropy.

It is worth mentioning that, despite this paper scenario is to contrast WhatsApp political and non-political public groups, we believe that our methodology is applicable to several scenarios and also to other instantaneous messaging platforms. Demonstrating this applicability is one of our future work directions, together with better understanding the role of each user in the group, as well as how it evolved across time. We also want to correlate observed behavior to external events and to assess how the impact of such events varies among groups. 

We expect that the findings provided in this paper contribute to shed light on the way WhatsApp works and reduce the opaqueness of the modern services of the global communication and information infrastructure.


\bibliographystyle{aaai}
\balance
\bibliography{00_Main}

\end{document}